# Component-Based Distributed Framework for Coherent and Real-Time Video Dehazing

Meihua Wang[1], Jiaming Mai[1], Yun Liang[1], Tom Z. J. Fu[2, 3], Zhenjie Zhang[3], Ruichu Cai[2]

[1]College of Mathematics and Informatics, South China Agricultural University, China

[2]School of Computer Science and Technology, Guangdong University of Technology, China

[3]Advanced Digital Sciences Center, Illinois at Singapore Pte.Ltd., Singapore

**Abstract.** Traditional dehazing techniques, as a well-studied topic in image processing, are now widely used to eliminate the haze effects from individual images. However, even the state-of-the-art dehazing algorithms may not provide sufficient support to video analytics, as a crucial pre-processing step for video-based decision making systems (e.g., robot navigation), due to the limitations of these algorithms on poor result coherence and low processing efficiency. This paper presents a new framework, particularly designed for video dehazing, to output coherent results in real time, with two novel techniques. Firstly, we decompose the dehazing algorithms into three generic components, namely transmission map estimator, atmospheric light estimator and haze-free image generator. They can be simultaneously processed by multiple threads in the distributed system, such that the processing efficiency is optimized by automatic CPU resource allocation based on the workloads. Secondly, a cross-frame normalization scheme is proposed to enhance the coherence among consecutive frames, by sharing the parameters of atmospheric light from consecutive frames in the distributed computation platform. The combination of these techniques enables our framework to generate highly consistent and accurate dehazing results in real-time, by using only 3 PCs connected by Ethernet.

## 1 Introduction

Video dehazing is an important module to video analytical systems, especially for video-based decision making applications, such as security surveillance and robot navigation. As an important pre-processing step, video dehazing is expected to recover the visual details of target objects in the video, even when the videos are recorded in extremely foggy environment. Inaccurate and inconsistent outputs of the dehazing component may decisively ruin the usefulness of the whole system, regardless of the performance of other video analytical modules in the rest of the system [1].

While conventional dehazing techniques are fairly mature in image processing, even the state-of-the-art image dehazing algorithms are not directly applicable to video dehazing [2]. This is mainly due to two major limitations on the design of these algorithms. Firstly, image dehazing is supposed to process one image at a time, without considering the coherence between computation results across frames in the video stream. In video dehazing, however, such coherence among consecutive frame is crucial, because the fluctuation on the contrast and color over the target objects could bring additional difficulties for further analysis, e.g. human tracking and detection. It is thus necessary to introduce normalization into dehazing algorithms to eliminate such undesirable variations. Secondly, most of the algorithms are basically too slow for real-time processing on video, therefore not suitable for real applications. Simple parallelization and redesign for new hardware (e.g., GPU) on these algorithms, may not scale up the processing efficiency, because the computation on one frame may depend on the results of other frames, when normalization solution mentioned above is incorporated [3].

In this paper, we present a new general framework for coherent and real-time video dehazing. By employing our framework, the video analytical system can easily transform an existing image dehazing algorithm into a distributed version, and automatically deploy the algorithm on a distributed platform (e.g., Amazon EC2) for real-time video processing. Moreover, the parallel computation and coherence enforcement are transparent to the programmer, in the sense that the system itself is responsible for computation resource management and result normalization to guarantee processing efficiency and coherence. Such performance guarantees are delivered by devising two novel techniques in our framework. Specifically, the task decomposition and parallelization technique decomposes a wide class of dehazing algorithms into three generic computation components, such that each component can be processed by multiple threads on the distributed platform in parallel. Furthermore, in order to avoid rapid model variation over the frames, an automatic state synchronization mechanism is employed to normalize the atmospheric light parameters across consecutive frames. A demo video is available at https://youtu.be/ZuflaEHp_RE.

### 1.1 Previous Works

**Image Dehazing** Dozens of image dehazing methods are proposed in the past few years [4]. Based on the related images in freeing haze, these methods can be divided into two major classes, namely single image approaches and multi-images approaches. Single image approaches usually utilize the statistic information (e.g., histogram-based methods [5-7]) or some assistant information (e.g., depth image [8, 9]) to estimate the dehazed images. Tan et al. [10] apply Markov Random Field (MRF) to maximize the local contrast to achieve haze-free images. Fattal et al. [11] recover contrast and details of fogged color images by separately analyzing the shading and transmission function in local regions. Dong et. al [12] employ the sparse priors to restore the color dehazed image. He et al. [13] design dark channel prior (DCP) model to estimate the thickness of haze and the atmospheric scattering model to produce dehazed images. A number of variants of DCP molel are proposed in the literature, such as [14-22]. In [23], Zhu et al. construct a linear model for scene depth estimation with color attenuation prior (CAP) and learned the parameters of the model by a supervised method. Multi-images approaches usually employ certain reference images with close relationship of hazed images to achieve dehazing. Narasimhan et al. [27] utilize the images taken under different weather conditions at the same scene to estimate the structure and depth of the scene for image dehazing. Treibitz et al. [24] use two frames instantly taken at a scene with different states of light-source polarize and propose an active polarization descattering method to restore images. All these multi-image approaches work only when the images are taken under strict constraints, which may not be available in practice.

However the multi-images based methods need reference images taken in the specific scene, they fail in utilizing the images from dynamic scene or taken by moving camera.

**Video Dehazing.** A handful of video dehazing approaches directly borrow image dehazing techniques to separately recover frames one by one. All of these approach suffer from the lack of temporal and spatial constraint to preserve coherence and poor processing efficiency, making them useless in real systems. To improve the efficiency, Lv et al. [3] use a cross-bilateral filter and the inversed haze model to do video dehazing based on GPU model. However, it cannot provide good temporal coherence across frames. Tarel et al. [25] propose a fast algorithm to restore low contrast and visibility of images but have difficulties when processing objects at similar color. Kim et al. [7] propose to use overlapped block filter to reduce computational complexity. To preserve the coherence to avoid the flicker of dehazed video, Zhang et al. [2] employ the guided filter [28] to compute the transmission maps in video dehazing and design an MRF model to improve the spatial and temporal coherence of the transmission maps. This method focuses on smoothing the coherence and cannot produce more accurate dehazed result than the guided filter. Li et al. [1] propose a stereo reconstruction for video dehazing (SRVD) by computing the scene depth and preserving the coherence between scene depth and fog transmission at neighboring points. This method produces favorable results in videos with translation, but cannot work in videos with rotation for its disability in reconstructing the depth.

**Fast stream processing**. The present image or video dehazing methods usually focus on improving the accuracy rather than the efficiency. However, real-time video dehazing is critical for some application such as robot navigation and video surveillance. Recently, some fast stream processing systems are proposed, including Apache Storm [29], Twitter Heron [30], StreamCloud [31] and Resa [32], to support generic distributed stream processing. Real-time applications, especially for video stream analytics, are built on top of these systems. Fu et al. [33] utilize Storm and DRS [37] to design a real-time trajectory tracking system over live videos. Zhang et al. [26] propose a storm-based framework to achieve real-time video object detection. Zhang et al. [34] combine Storm stream processing and Hadoop batch processing to reveal knowledge hidden in video data by deep learning. These methods have demonstrate the possibilities of video stream processing on distributed platforms.

## 1.2 Contributions

The goal of the paper is to present a new framework to generate high-quality dehazed video in real-time, by exploiting the computation power of distributed computing. To achieve this goal, we design a component-based framework to coordinate the estimators of haze-free video. In particular, the framework is friendly to distributed computing, in the sense that each component could be easily paralleled and runs on different frames at the same time. As a short summary, our main contributions are listed below:

- A decomposition model is designed to decompose existing image or video dehazing approaches into three generic components based on the basic physical model of hazy video formation.
- A new atmospheric light estimator update strategy is proposed to normalize the atmospheric light values among consecutive frames in the video, preserving the spatial-temporal coherence and avoiding potential flickers on the outputs.
- A distributed processing scheme is built to simultaneously running components for different frames to generate the dehiring results in real time.

## 2 Background

The physical model, named as atmospheric scattering model [1], has been widely used in image and video dehazing for its excellent performance on hazy images modeling. Basically, it is based on the following mathematical model:

$$\mathbf{I}(x) = \mathbf{J}(x)t(x) + \mathbf{A}(1-t(x)), \tag{1}$$

where $x$ is a 2-dimensional coordinate of the pixel within the image, $\mathbf{I}$ is the hazy image, and $\mathbf{A}$ is the atmospheric light, i.e., a 3-dimensional RGB vector, $\mathbf{J}$ denotes the scene radiance, which is also the haze-free image we try to generate, and *t* is the transmission map considered as the thickness of the hazes. Here, *t* can be further determined by the distance between the scene and the observer:

$$t(x) = e^{-\beta d(x)}, \qquad (2)$$

where *β* is the scattering coefficient, and *d* is the scene depth map. Intuitively, *t(x)*=0 means that it is completely hazy while *t(x)*=1 means haze-free.

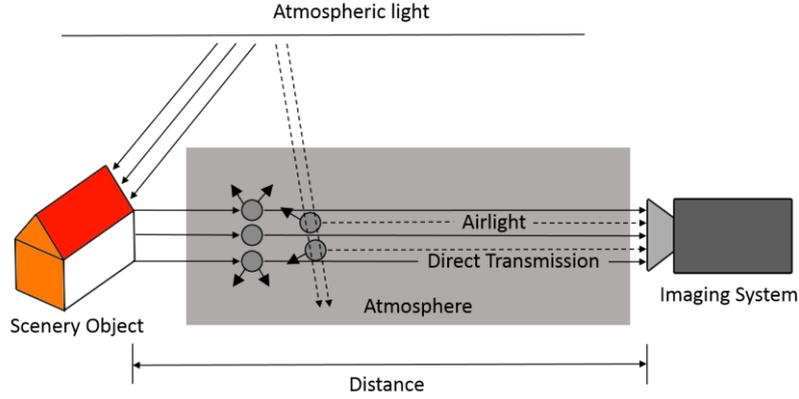

**Fig. 1.** Atmospheric scattering model.

The target of dehazing is to restore **J** given **I**. As *t*, **A** and **J** are unknown variables in Equation 1, dehazing is an ill-posed problem. We need to estimate *t* and **A** first, which are used to restore the haze-free image **J**. Equation 1 and Equation 2 have modeled two mechanisms of imaging in hazy environment. We illustrate the two mechanisms in Figure 1. The first mechanism occurs on direct transmission and describes the attenuation of light when it travels through the atmosphere. It is modeled by **J**(*x*)*t*(*x*) in Equation 1, and is a multiplicative effect. The second mechanism is the airlight which is caused by the scattering of environmental illumination. It is modeled by **A**(1-*t(x)*) in Equation 1 and introduces additive influence in hazy formation.

## 3   Our Proposed Framework

This section proposes our component-based distributed framework for video dehazing. This framework is formulated based on the dehazing components decomposition, component-based distributed computing network, and atmospheric light update strategy. With the decomposition, we decompose a dehazing algorithm into three independent components. Given the distributed computation resource, the framework parallelize the components to achieve real-time video dehazing. By including the update strategy, the framework also preserves the coherence between frames to eliminate potential flicker in the result video.

### 3.1   Dehazing Components Decomposition

Based on the physical model described above, the common workflow of dehazing approaches is the calculation on transmission map and atmospheric light, which are used to restore hazed images. By abstracting the workflow without the detailed algorithm on the common subproblems, we decompose a wide class of dehazing algorithms into three components. The first component is a transmission map estimator which is used to compute the *t* in Equation 1. The second component is an atmospheric light estimator which is used to calculate the **A** in Equation 1. The third component is a haze-free image generator estimating *t* and **A** to generate the haze-free image. The overall structure of the framework with the components is depicted in Figure 2.

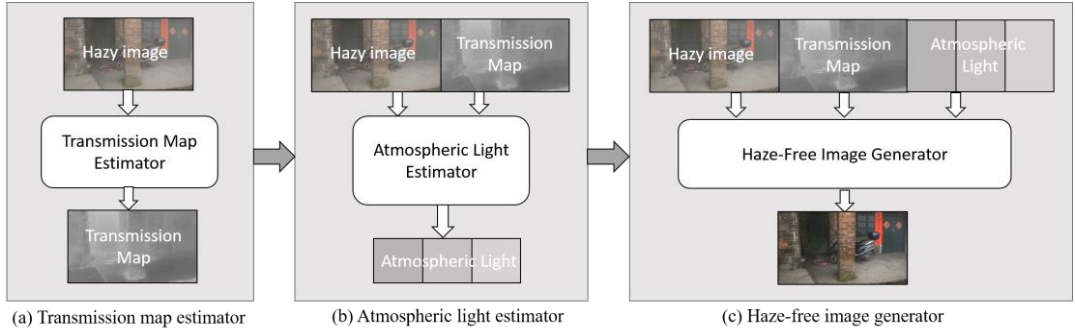

(a) Transmission map estimator  (b) Atmospheric light estimator  (c) Haze-free image generator

**Fig. 2.** The three components of a dehazing algorithm.

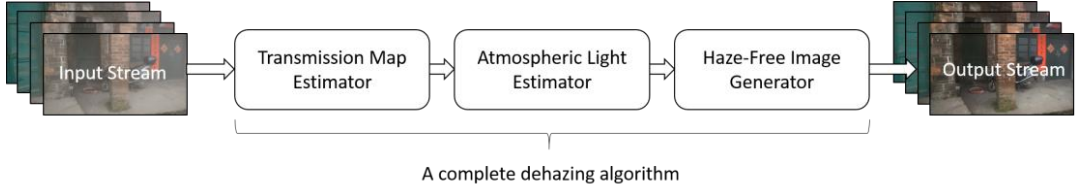

**Fig. 3.** The flow of video dehazing based on three components.

**Transmission Map Estimator**: It is defined to compute the transmission map by inputting a hazed image into a dehazing method. As we focus on designing a framework general to each dehazing method, the detail of how the compute the transmission map is a black box. But it turns to be the specific algorithm when a dehazing method is projected on this frame. Here we take two famous image dehazing methods, the DCP [13] and the CAP [23] as examples for this estimator.

The DCP method estimates the transmission *t* of pixel *x* based on color channel $I^c$ of hazy image **I**, and atmospheric light of color channel $A^c$ by:

$$t(x) = 1 - \min_{y \in \Omega(x)} (\min_c \frac{I^c(y)}{A^c}). \tag{3}$$

Similarly, the CAP method calculate the transmission based on the linear coefficients $\omega_0$, $\omega_1$ and $\omega_2$, the value channel *v* and saturation channel *s* by:

$$t(x) = e^{-\beta(\omega_0 + \omega_1 v(x) + \omega_2 s(x))}, \tag{4}$$

The transmission map estimator is the first process of dehazing, its output provides the basic preprocess data for the atmospheric light estimation. As the overview of image dehazing process in Figure 3, the estimated transmission map is the essential input of atmospheric light estimator.

**Atmospheric Light Estimator**: For the physical model described in Equation 1 is general for all dehazing methods, a common method to design atmospheric light estimator is extracted from Equation 1. When *t* tends to zero, Equation 1 degenerates to **A**=**I**(*x*). This shows that **A** can be estimated by **I**(*x*) at pixel *x* which makes *t*(*x*) small enough. Therefore, we compute **A** by:

$$\mathbf{A} = \mathbf{I}(x), t(x) < t_{threshold}. \tag{5}$$

Further, assuming that the smallest value of the transmission map *t* is even smaller than the $t_{threshold}$, **A** is finally evaluated by:

$$\mathbf{A} = \mathbf{I}(x), x \in \{x \mid \forall y : t(y) > t(x)\}. \tag{6}$$

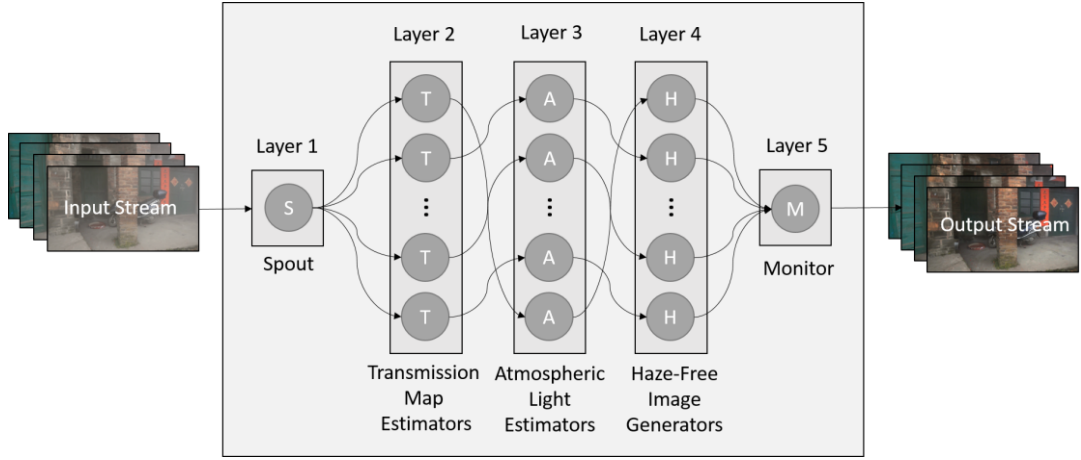

**Fig. 4.** The overview of component-based distributed framework for video dehazing.

Given the common model described in Equation 7, the atmospheric light estimator utilizes a hazy image and its estimated transmission map as input to compute atmospheric light **A** in Equation 1.

As shown in Figure 3, this atmospheric light estimation is used to solve Equation 1 to calculate haze-free images by the following haze-free image generator.

**Haze-Free Image Generator**: It generates the haze-free image **J** with the transmission map $t$ and the atmospheric light **A** which are estimated by the estimators mentioned above. From Equation 1, we compute **J** by:

$$\mathbf{J}(x) = \frac{\mathbf{I}(x) - \mathbf{A}}{t(x)} + \mathbf{A}. \tag{7}$$

To avoid too much noise, $t(x)$ is usually restricted by a lower bound $t_0$=0.1:

$$\mathbf{J}(x) = \frac{\mathbf{I}(x) - \mathbf{A}}{\max\{t(x), t_0\}} + \mathbf{A}. \tag{8}$$

Similar to the atmospheric light estimator, the haze-free image generator is still common and generally used in exiting dehazing methods. It is the last component to produce haze-free images. As described in Figure 3, for a process of dehazing an image, we need to orderly implement the three components. Therefore, if only dehazing an image, there is no parallel processing in component implementation. In fact, the high time cost of video dehazing occurs on frame by frame processing. To tackle the efficiency issue, we design a component-based distributed computing network to parallelize the components on consecutive frames to achieve real-time video dehazing.

### 3.2   Component-Based Distributed Computing Network

Three reasons inspire us to construct this distributed computing network to achieve video dehazing. The first is the real-time processing requirement and the high time cost of consecutive dehazing under frame by frame. The second is that the component separation of dehazing method makes the distributed dehazing become possible. The third one is that the development of fast data processing introduces us how to design the network to provide distributed implementation.

As described in Figure 4, our network is constructed by five layers. The first layer is defined by a spout which is used to receive the input video stream and distribute them to the spare transmission map estimator in the second layer. The second layer is formed by the transmission map estimators. Each of them waits for processing the hazy frame image from the spout, then outputs the estimated transmission map to the spare atmospheric light estimator in the third layer. Similarly, each atmospheric light estimator computes the atmospheric light with a transmission map and its hazy image, and pass it on to the next layer. In the forth layer, the spare haze-free image generator receives the original hazy image, the transmission map, and the atmospheric light as paramters to produce a haze-free image. Finally, in the last layer, the monitor is defined to sort the generated haze-free images and display (or write) the output stream.

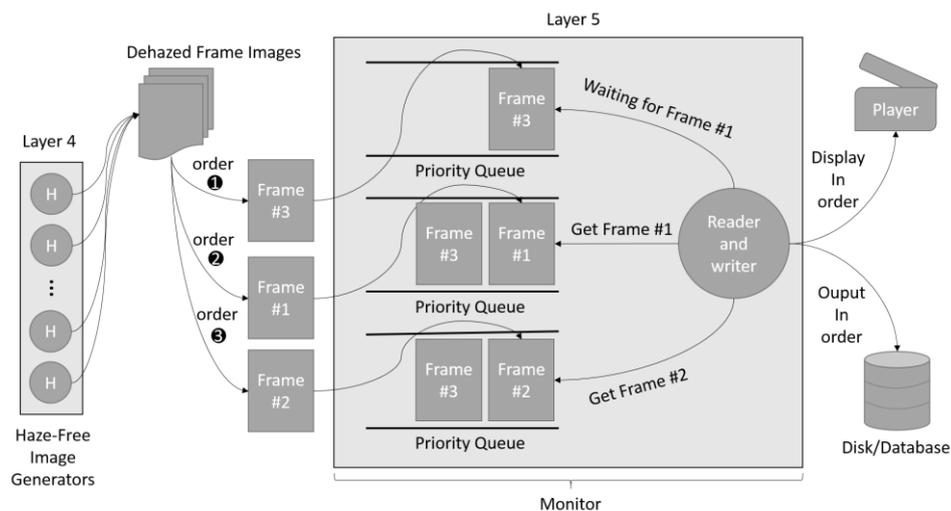

**Fig. 5.** The working principle of the monitor component in the fifth layer.

The proposed network achieves parallel process by designing several nodes for each component. As shown in Figure 4, if there are *N* nodes for transmission map estimator, we can parallel compute *N* transmission maps for *N* frames. Here each node corresponds to an operator which is related to a thread in our network. Many threads can be projected to more than one PC to share computation. Therefore, the order of dehazed frames generated in the fourth layer is not guaranteed (i.e. the third dehazed frame may be generated before the first dehazed frame). Therefore, in the last layer of network, we design a monitor to keep dehazed video formed in right order. Figure 5 describes the construction of the monitor which includes a priority queue, a reader and a writer. The monitor performs as follow. First, the dehazed frames generated by the forth layer is putted into the priority queue constantly. Then the priority queue sorts them according to their frame ID. Meanwhile, the reader extracts frames in the queue sequentially. If the reader meets an absent frame during consecutive reading, it will wait for a period (e.g. 20 milliseconds). To keep real-time processing, the reader will skip that frame and read the next one after timeout. Finally, the writer displays the dehazed frames in order, or output them to the disk (or database).

### 3.3 Atmospheric Light Update Strategy

In the sequence of a video, atmospheric light change is very popular. To update it is very essential to preserve the quality of dehazed images. The poor update leads to unsmoothing change in atmospheric light which finally introduces flicker. Taking Figure 6 as an example, in the second row, the luminance between adjacent frames is unsmooth, and flicker occurs. Here, the 191th frame is darker than 190th frame while much brighter than the 190th and 192th frame. This is because there are subtle errors in atmospheric light estimation. The third row provides better results by coordinating the estimated atmospheric light.

Most of the present dehazing methods updates the atmospheric light by independently computing it for each frame. It will cost much time and cannot preserve the coherence between frames. In our framework, we define a new method to update atmospheric light based on the parallel processing. As described in [1], the adjacent frames in a video share many features. It is reasonable that the adjacent frames own similar atmospheric light. Therefore, we only update the atmospheric light for every several frames.

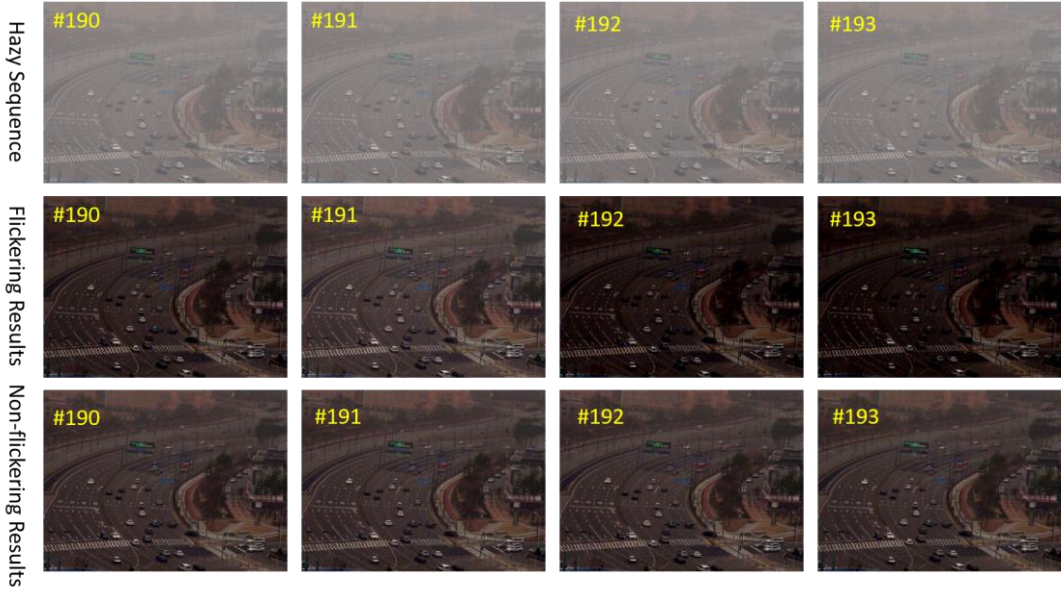

**Fig. 6.** Flicker caused by estimating the atmospheric light every frame independently.

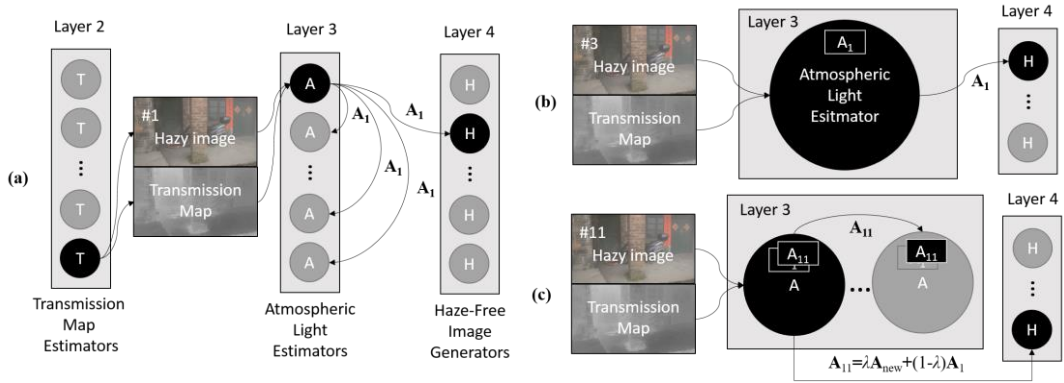

**Fig. 7.** The atmospheric light update strategy. (a) the estimator computes the atmospheric light and sends it to other estimators as well as the spare haze-free image generator. (b) the estimator omits computation of the atmospheric light if the distance between the received frame and the saved frame is small. (c) estimate and update the atmospheric light if this distance is large enough.

The construction of our update scheme is described in Figure 7. Firstly, a spare atmospheric light estimator in the third layer, is selected to estimate the atmospheric light and send the estimated result to all other estimators and the haze-free image generator in the next layer. The haze-free image generator uses it to produce the dehazed image, while other atmospheric light estimators save it independently. As described in Figure 7 (a), The black "A" in the third layer is the selected estimator which estimates the atmospheric light of frame 1 and sends it to the related estimators. Now setting a value $l$ which is used for describing the update frequency, all the estimators will omit computation of the atmospheric light, if the distance between the received frame $m$ and the saved frame $k$ is too small (i.e. $m-k<l$). This situation is described in Figure 7(b), namely as the distance between frame 3 and frame 1 small enough, it reuses the estimated atmospheric light of frame 1 for omitting computation. On the other hand, if the difference is large enough (i.e. $m-k>=l$), the estimator estimates and updates the atmospheric light by the following equation:

$$\mathbf{A}_m = \lambda \mathbf{A}_{new} + (1-\lambda)\mathbf{A}_k, \qquad (9)$$

where $\mathbf{A}_{new}$ is the estimated atmospheric light, $\lambda$ is a scalar parameter in [0, 1], and $\mathbf{A}_m$ is the atmospheric light of the $m$th frame. In general, we set $\lambda=0.05$ for guaranteeing that the atmospheric light varies smoothly. As shown in Figure 7(c), we finally update all the atmospheric light estimators and overwrite it from $\mathbf{A}_k$ to $\mathbf{A}_m$. Meanwhile, $\mathbf{A}_m$, rather than $\mathbf{A}_k$, will be sent to the spare haze-free image generator in the next layer for haze removal.

With the proposed update strategy, we reduce the errors of atmospheric light introduced by the noise of adjacent frames and provide smooth estimated values. Therefore, our method can avoid the flickering problem in dehazed images. Another advantage is that it reduces the time cost compared with the original frame by frame updating method. As demonstrated in Figure 6, the third row is computed by our update strategy and performs better with stable luminance than the second row which is achieved by frame by frame updating.

## 4  Experiments

In this section, we present the details of our implementation (Section 4.1), and evaluate the efficiency and quality of our approach against state-of-the-art solution in the literature (Section 4.2).

### 4.1  Implementation Details

We implement the proposed framework using Apache Storm [29]. In qualitative evaluation, we carry out the experiments on a cluster with 4 nodes. One of the nodes is used as the master node (Zookeeper and Nimbus), and other three nodes are slave nodes. All the nodes are equipped with a 2.50GHz CPU and 16GB RAM. Each node runs 3 threads concurrently, which sums up to 12 threads in total. We implement the state-of-the-art dehazing algorithms with the proposed framework, and compare their results to those of the original version. In quantitative evaluation, we analyze the efficiency improvement of our framework. The parameters investigated in the experiments include the resolution of the input video and the number of nodes used in the computation. In each experiment, we vary one parameter and fix all others.

### 4.2  Efficiency Evaluation

Table 1 shows the average number of frames processed by DCP [13], CAP [23], SRVD [1], and the distributed version of DCP and CAP. 1N-DCP indicates that only 1 node is used for processing of DCP on the distributed environment. Similarly, 2N-DCP and 3N-DCP employs 2 nodes and 3 nodes, respectively. Three different resolutions of the videos are tested in our experiments. For each resolution, we test on 20 videos with average results reported in the table.

As the results show, in terms of the 320×240's videos, the proposed framework significantly improves the frame rate of the DCP, above 150 frames per second, nearly 7 times faster than conventional single-machine implementation. Under the same conditions, the efficiency of CAP is improved by 5 times. With 3 working nodes, our framework can process more than 250 frames per second. In the case of the resolution of 640×480, both the DCP and the CAP rapidly slow down. The frame rates are only 5.29 frames/s and 10.03 frames, respectively. By using our proposed framework, the rates grow to 35.08 frames/s and 56.21 frames/s, which is more than enough for practical real-time video dehazing. The speed of the two approaches is even slower when processing videos with a resolution of 1024×576. Both of them can only process less than 5 frames every second. This is far from satisfactory, as high-resolution videos are becoming popular. By extending the number of nodes on the cluster environment, they are sufficiently efficient to meet real-time response time requirement (29.46 frames/s with 3N-DCP and 37.55 frames/s with 3N-CAP). In contrast, the SRVD is much slower, since it includes the computational intensive matting operation [36]3 which emphasizes the dehazing quality at the expense of efficiency.

Table 1. Time consumption comparison with different dehazing algorithm

| Algorithms | 320×240's videos | 640×480's videos | 1024×576's videos |
| --- | --- | --- | --- |
| DCP [13] | 23.17 frames/s | 5.29 frames/s | 2.72 frames/s |
| 1N-DCP | 60.60 frames/s | 17.24 frames/s | 9.63 frames/s |
| 2N-DCP | 121.28 frames/s | 27.20 frames/s | 17.38 frames/s |
| 3N-DCP | 150.02 frames/s | 35.08 frames/s | 29.46 frames/s |
| CAP [23] | 43.66 frames/s | 10.03 frames/s | 4.89 frames/s |
| 1N-CAP | 99.32 frames/s | 38.07 frames/s | 19.10 frames/s |
| 2N-CAP | 195.79 frames/s | 45.45 frames/s | 28.23 frames/s |
| 3N-CAP | 266.56 frames/s | 56.21 frames/s | 37.55 frames/s |
| SRVD [1] | 0.0059 frames/s | 0.0016 frames/s | 0.00075 frames/s |

Notice the fact that due to the elasticity of the framework, adding more nodes are useful for further improving the ability to deal with frames in batch. As shown in Table 1, 2N-DCP gets a higher speed compared with 1N-DCP in all conditions, while 3N-DCP perform even better. Similar conclusion can be found in 1N-CAP, 2N-CAP and 3N-CAP.

### 4.3 Quality Evaluation

In Figure 8, we give the curve graphs of the estimated atmospheric light on four different videos. Figure 8(a-d) are four different videos. There are two curve graphs for each video. One is for the DCP [13], the other is for the CAP [23]. In each curve graph, we use red, light green and blue to demonstrate the RGB value of the atmospheric light estimated by the original approaches of DCP and CAP. Then pink, dark green and cyan are used for the atmospheric light that is estimated by the update strategy mentioned in section 3.3. By analyzing the curve graphs, it is not difficult to find that oscillation appears in the curve, meaning the atmospheric light estimated by the original approaches varies frequently. This leads to serious flicker during dehazing as described in section 3.3. In contrast, that estimated by the update strategy keeps stable and varies smoothly. This contributes to solve the flickering problem in video dehazing.

In Figure 9, we show the results comparison of different dehazing algorithms. The first column is the haze frames from different videos. The second to the forth columns are the results of DCP, CAP, and SRVD, respectively. The results of DCP and CAP which are implemented by the proposed framework are shown in the fifth and the sixth columns. It is observed that the dehazing quality is well preserved by our framework. In addition, we improve the dehazing methods by providing more reasonable luminance than the present methods. The luminance of the results conducted by the framework are even much more similar to that of the realistic scene. This is due to the effectiveness and efficiency of the proposed update strategy for atmospheric light estimation. The proposed update strategy provides a new method to preserve the spacial and temporal coherence of video dehazing.

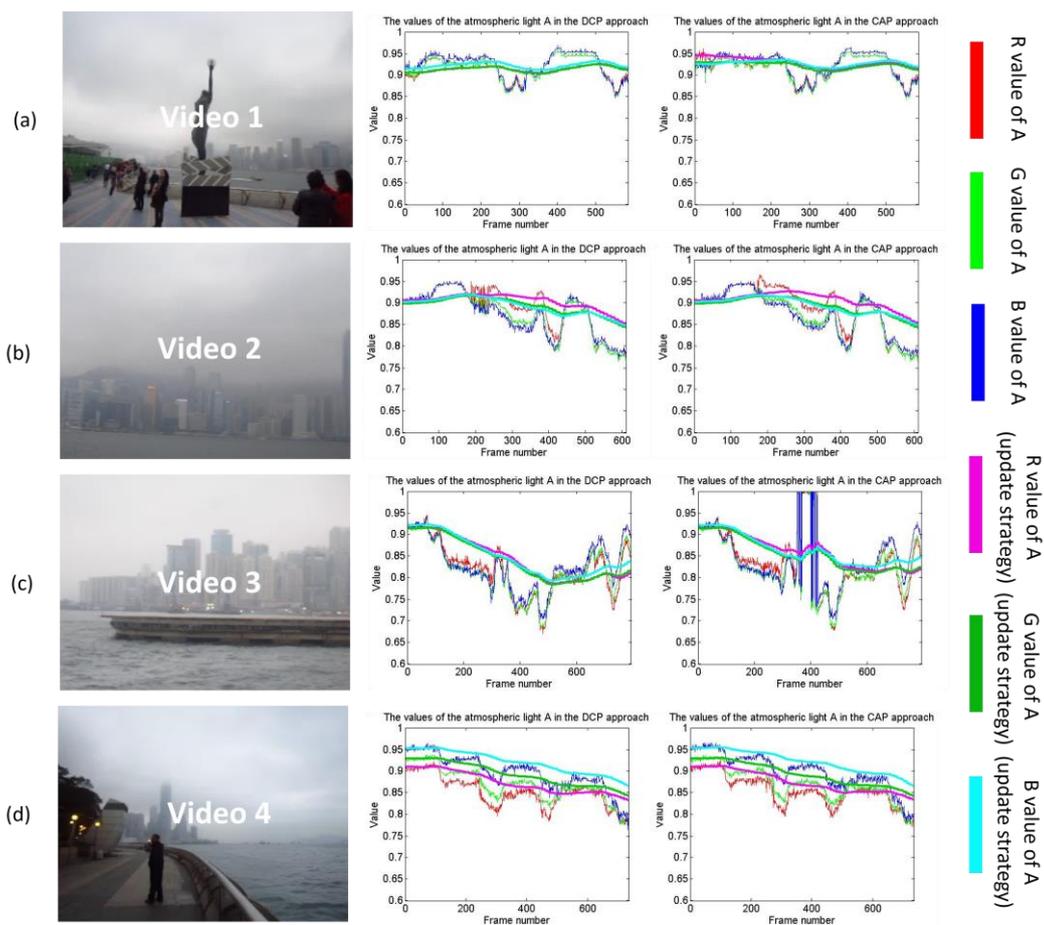

**Fig. 8.** Curve graphs of the estimated atmospheric light on different videos.

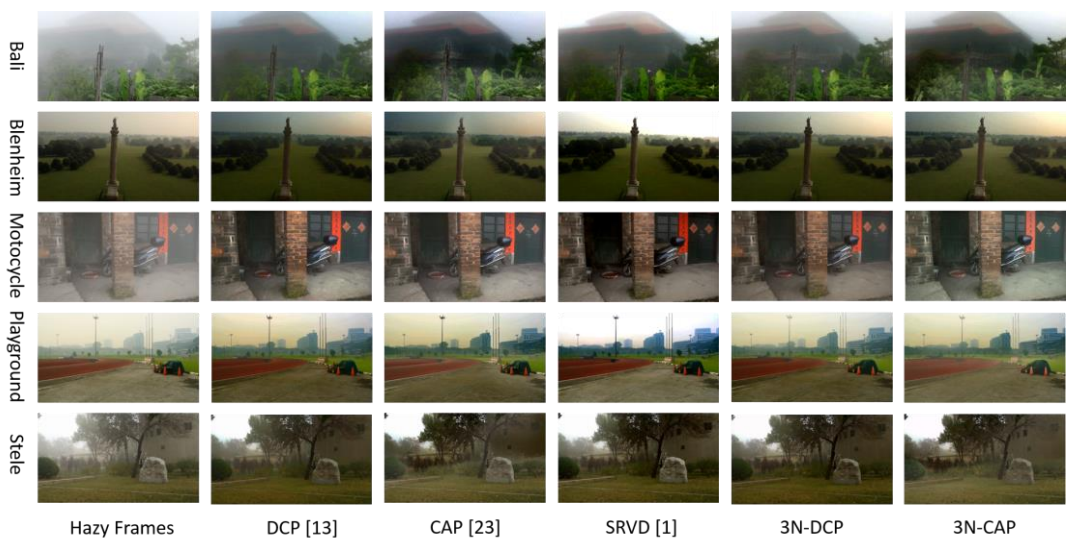

**Fig. 9.** Comparisons of deazhing results on videos..

## 5  Conclusion and Future Work

This paper proposes a component-based distributed framework to achieve real-time processing and preserve spatial-temporal coherence of video dehazing. Our framework is formed by three parts, namely dehazing components decomposition, component-based distributed computing network and atmospheric light update strategy. By the components decomposition, every exiting image or video dehazing methods can be projected onto this framework. The decomposed components are treated as separate operators and implemented by our proposed computing network. This network provides the parallel processing of the components, and automatically scales up and down based on the actual workload. The proposed updating scheme provides a new method to estimate the atmospheric light which performs favorable in preserving spatial-temporal coherence of video dehazing. Experiments show our framework easily scales up the efficiency of state-of-the-art solution to meet real-time requirement. One possible future work is to simultaneously process a number of videos at the same time, such that the atmospheric light estimation can be shared and coordinated between different videos.

## 6  Acknowledgement

This work was supported by the National Science Fund of China (U1501254, 61202293) and the the Science and Technology Planning Project of Guang-dong Province (2014A050503057, 2016A020210087). Fu and Zhang are supported by the research grant for the HCCS Programme at the Advanced Digital Sciences Center from Singapore's Agency for Science, Technology and Research (A*STAR). Fu and Zhang are also supported by Science and Technology Planning Project of Guangdong under grant (No. 2015B010131015).